\documentclass[12pt,tightenlines]{revtex4}
\usepackage{graphicx}
\usepackage{epsfig}
\def\etal{{\it et al.}}
\renewcommand\prd[3]{{\it Phys.\ Rev.\ }{\bf D #1}, #2 (#3)}
\newcommand\plb[3]{{\it Phys.\ Lett.\ }{\bf B #1}, #2 (#3)}
\renewcommand\prl[3]{{\it Phys.\ Rev.\ Lett.\ }{\bf #1}, #2 (#3)}

\begin{document}
\bibliographystyle{revtex}
\vspace{10mm}
\title{Probing New Physics With {\boldmath $b$} Decays} % actual title

\author{Sheldon Stone}
\email[]{Stone@physics.syr.edu}
\affiliation{\it Physics Dept., Syracuse University, USA, 13244-1130} 

%\date{\today}

\begin{abstract}
I discuss how $b$ decays can be used to unravel new physics beyond the Standard
Model. Decays second order in the weak interaction involving loops and CP
violation are emphasized. This information is complementary to
that obtainable with higher energy machines.
\end{abstract}
% \pacs{}
\maketitle

\vspace{-2.7in}
\vbox{\rightline{\hfil SUHEP 2001-10}
\rightline{\hfil Nov. 2001~~~~~~~~}}
\vspace{2in}
\vspace{6in}
\begin{flushleft}
.\dotfill .
\end{flushleft}
Presented at Snowmass 2001, Working Group on Flavor Physics, P2, Snowmass, CO,
July, 2001.
\newpage
\section{Introduction}

There are many reasons why we believe that the Standard Model is incomplete and
there must be physics beyond. One is the plethora of ``fundamental parameters,"
for example quark masses, mixing angles, etc... The Standard Model cannot
explain the smallness of the weak scale compared to the GUT or Planck scales;
this is often called ``the hierarchy problem." It is
believed that the CKM source of CP violation in the Standard Model is not
large enough to explain the baryon asymmetry of the Universe \cite{Gavela};
we can also take the view 
that we will discover additional large unexpected effects in $b$ and/or
$c$ decays. Finally, gravity is not incorporated. John Ellis said ``My personal
interest in CP violation is driven by the search for physics beyond the
Standard Model" \cite{Ellis}.

We must realize that {\it all} our current measurements are a combination of
Standard Model and New Physics; any proposed models must satisfy current
constraints. Since the Standard Model tree level diagrams are probably large,
lets consider them a background to New Physics. Therefore loop diagrams and CP
violation are the best places to see New Physics. 
The most important current constraints on New Physics models are
\begin{itemize}
\item The neutron electric dipole moment, $d_N~<6.3\times 10^{-26}$e-cm.
\item ${\cal{B}}(b\to s\gamma)=(3.23\pm 0.42)\times 10^{-4}$ and
${\cal{B}}(b\to s\ell^+\ell^-)<4.2\times 10^{-5}$.
\item CP violation in $K_L$ decay, $\epsilon_K =(2.271\pm 0.017)\times
10^{-3}$.
\item $B^o$ mixing parameter $\Delta m_d = (0.487\pm0.014)$ ps$^{-1}$.
\end{itemize}

\section{Generic Tests for New Physics}
We can look for New Physics either in the context of specific models or more
generically, for deviations from the Standard Model expectation.

One example is to examine the rare decays $B\to K\ell^+\ell^-$ and
$B\to K^*\ell^+\ell^-$ for branching ratios and polarizations.
According to Greub et al. \cite{Greub95}, ``Especially the decay into $K^*$ yields a wealth
of new information on the form of the new interactions since the Dalitz
plot is sensitive to subtle interference effects."

\begin{figure}[htb]
\vspace{0.4cm}
\centerline{\epsfig{figure=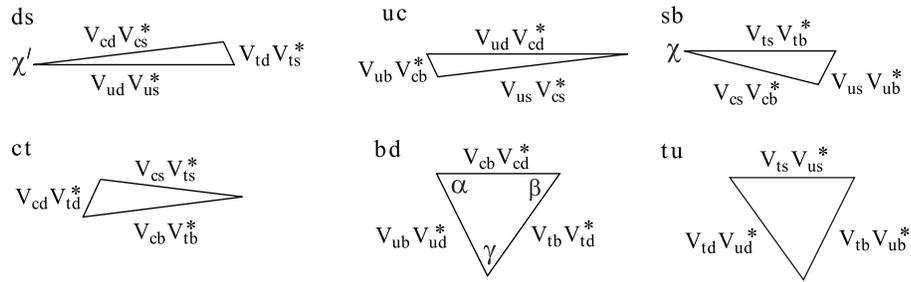,height=1.5in}}
%\vspace{-1.1cm}
\caption{\label{six_tri}The six CKM triangles. The bold labels, e.g. {\bf ds} 
refer to the rows or columns used in the unitarity relationship. The angles
defined in equation (\ref{eq:chi}) are also shown.}
\end{figure}

Another important tactic is to test for inconsistencies in Standard Model
predictions independent of specific non-standard models.
 The unitarity of the CKM matrix allows us to construct six relationships.
These may be thought of as triangles in the complex plane  
shown in Fig.~\ref{six_tri}. 

All six of these triangles can be constructed knowing four and
only four independent angles \cite{silva_wolf}\cite{KAL}\cite{bigis}.
 These 
are defined as:
\begin{eqnarray} \label{eq:chi}
\beta=arg\left(-{V_{tb}V^*_{td}\over V_{cb}V^*_{cd}}\right),&~~~~~&
\gamma=arg\left(-{{V^*_{ub}V_{ud}}\over {V^*_{cb}V_{cd}}}\right), \\
\chi=arg\left(-{V^*_{cs}V_{cb}\over V^*_{ts}V_{tb}}\right),&~~~~~&
\chi'=arg\left(-{{V^*_{ud}V_{us}}\over {V^*_{cd}V_{cs}}}\right).\nonumber\\ 
\end{eqnarray}
($\alpha$ can be used instead of $\gamma$ or $\beta$.) Two of the phases $\beta$ and $\gamma$ are probably large while $\chi$ is
estimated to be small $\approx$0.02, but measurable, while $\chi'$ is likely
to be much smaller.

It has been pointed out by Silva and Wolfenstein \cite{silva_wolf} that
measuring only angles may not be sufficient to detect
new physics. For example, suppose there is new physics that arises in 
$B^o-\overline{B}^o$ mixing. Let us assign a phase $\theta$ to this new
physics. If we then measure CP violation in $B^o\to J/\psi K_s$ and eliminate
any Penguin pollution problems in using $B^o\to\pi^+\pi^-$, then we actually
measure $2\beta' =2\beta + \theta$ and $2\alpha' = 2\alpha -\theta$. So while
there is new physics, we miss it, because
$2\beta' + 2\alpha' = 2\alpha +2\beta$ and $\alpha' + \beta' +\gamma
= 180^{\circ}$.

\subsection{A Critical Check Using $\chi$}

The angle $\chi$, defined in equation~\ref{eq:chi}, can be extracted by
measuring the time dependent CP violating asymmetry in the reaction
$B_s\to J/\psi \eta^{(}$$'^{)}$, or if one's detector is incapable of quality
photon detection, the $J/\psi\phi$ final state can be used.  However, in this
case there are
two vector particles in the final state, making this a state of mixed CP,
requiring a time-dependent angular analysis to extract $\chi$, that requires large
statistics.

Measurements of the magnitudes of 
CKM matrix elements all come with theoretical errors. Some of these are hard 
to estimate.
The best measured magnitude is that of $\lambda=|V_{us}/V_{ud}|=0.2205\pm 
0.0018$. 
Silva and 
Wolfenstein \cite{silva_wolf} \cite{KAL}
show that the Standard Model 
can be checked in a profound manner by seeing if:
\begin{equation}
\sin\chi = \left|{V_{us}\over 
V_{ud}}\right|^2{{\sin\beta~\sin\gamma}\over{\sin(\beta+\gamma)}}~~.
\end{equation}
Here the precision of the check will be limited initially by the measurement of
$\sin\chi$, not of $\lambda$. This check can  reveal new physics, even 
if other measurements have not shown any anomalies. 
Other relationships to check include:
\begin{eqnarray}
\sin\chi = \left|{V_{ub}\over 
V_{cb}}\right|^2{\sin\gamma~\sin(\beta+\gamma)\over{\sin\beta}}~~,
&~~~~~~&
\sin\chi = \left|{V_{td}\over 
V_{ts}}\right|^2{\sin\beta~\sin(\beta+\gamma)\over{\sin\gamma}}~~.
\end{eqnarray}

The astute reader will have noticed that these two equations lead to the
non-trivial relationship:
\begin{equation}
\sin^2\beta\left|V_{td}\over V_{ts}\right|^2 = \sin^2\gamma\left|{V_{ub}\over 
V_{cb}}\right|^2 ~~.
\end{equation}
This constrains these two magnitudes in terms of two of the angles. 
Note, that it is in principle possible to determine the magnitudes of 
$|V_{ub}/V_{cb}|$ and $|V_{td}/V_{ts}|$ without model dependent errors
by measuring
$\beta$, $\gamma$ and $\chi$ accurately. Alternatively, $\beta$, $\gamma$
 and $\lambda$ can be used to give a much more precise value than is
possible at present with direct methods. For example, once $\beta$ and
$\gamma$ are known
$\left|{V_{ub}/ V_{cb}}\right|^2 = \lambda^2
\sin^2\beta/ {\sin^2(\beta +\gamma)}$.

Table~\ref{table:reqmeas} lists the most important physics quantities and the
decay modes that can be used to measure them. The necessary detector
capabilities include the ability to collect purely hadronic final states
labeled here as ``Hadron Trigger," the ability to identify
charged hadrons labeled as ``$K\pi$ sep," the ability to detect photons with
good efficiency and
resolution and excellent time resolution required to analyze rapid $B_s$
oscillations. Measurements of $\cos(2\phi)$ can eliminate 2 of the 4 ambiguities in $\phi$ that are present when $\sin(2\phi)$ is measured. 

\begin{table}[hbt]
\begin{center}
\caption{Required CKM Measurements for $b$'s}
\label{table:reqmeas}
%\vspace*{2mm}
\begin{tabular}{|l|l|c|c|c|c|} \hline\hline
Physics & Decay Mode & Hadron & $K\pi$ & $\gamma$ & Decay \\
Quantity&            & Trigger & sep   & det & time $\sigma$ \\
\hline
$\sin(2\alpha)$ & $B^o\to\rho\pi\to\pi^+\pi^-\pi^o$ & $\surd$ & $\surd$& $\surd$ 
&\\
$\cos(2\alpha)$ & $B^o\to\rho\pi\to\pi^+\pi^-\pi^o$ & $\surd$ & $\surd$& 
$\surd$ &\\
sign$(\sin(2\alpha))$ & $B^o\to\rho\pi$ \& $B^o\to\pi^+\pi^-$ & 
$\surd$ & $\surd$ & $\surd$ & \\
$\sin(\gamma)$ & $B_s\to D_s^{\pm}K^{\mp}$ & $\surd$ & $\surd$ & & $\surd$\\
$\sin(\gamma)$ & $B^-\to \overline{D}^{0}K^{-}$ & $\surd$ & $\surd$ & & \\
$\sin(\gamma)$ & $B^o\to\pi^+\pi^-$ \& $B_s\to K^+K^-$ & $\surd$ & $\surd$& & 
$\surd$ \\
$\sin(2\chi)$ & $B_s\to J/\psi\eta',$ $J/\psi\eta$ & & &$\surd$ &$\surd$\\
$\sin(2\beta)$ & $B^o\to J/\psi K_s$ & & & & \\
$\cos(2\beta)$ &  $B^o\to J/\psi K^o$, $K^o\to \pi\ell\nu$  & &$\surd$ & & \\
$\cos(2\beta)$ &  $B^o\to J/\psi K^{*o}$ \& $B_s\to J/\psi\phi$  & & & 
&\\
$x_s$  & $B_s\to D_s^+\pi^-$ & $\surd$ & & &$\surd$\\
$\Delta\Gamma$ for $B_s$ & $B_s\to  J/\psi\eta'$, $ D_s^+\pi^-$, $K^+K^-$ &
$\surd$ & $\surd$ & $\surd$ & $\surd$ \\
\hline
\end{tabular}
\end{center}
\end{table}

\subsection{Finding Inconsistencies}
Another interesting way of viewing the physics was given by Peskin
\cite{Peskin}. Non-Standard Model physics would show up as discrepancies among
the values of $(\rho ,\eta)$ derived from independent determinations using CKM
magnitudes ($|V_{ub}/V_{cb}|$ and $|V_{td}/V_{ts}|$), or $B^o_d$ mixing ($\beta$ and
$\alpha$), or $B_s$ mixing ($\chi$ and $\gamma$).

\subsection{Required Measurements Involving $\beta$}

Besides a more precise measurement of $\sin 2\beta$ we need to 
resolve the ambiguities. There are two suggestions on how
this may be accomplished. Kayser \cite{kkayser} shows that time dependent
measurements of the final state
$J/\psi K^o$, where $K^o\to \pi \ell \nu$, give a direct measurement of
$\cos(2\beta)$ and can also be used for CPT tests. Another suggestion is to use
the final state $J/\psi K^{*o}$, $K^{*o}\to K_S\pi^o$, and to compare with
$B_s\to J/\psi\phi$ to extract the sign of the strong interaction phase shift
assuming SU(3) symmetry, and thus determine $\cos(2\beta)$ \cite{isi_beta}.

\subsection{Required Measurements Involving $\alpha$ and $\gamma$}
\begin{figure}[htb]
\vspace{-.4cm}
\centerline{\epsfig{figure=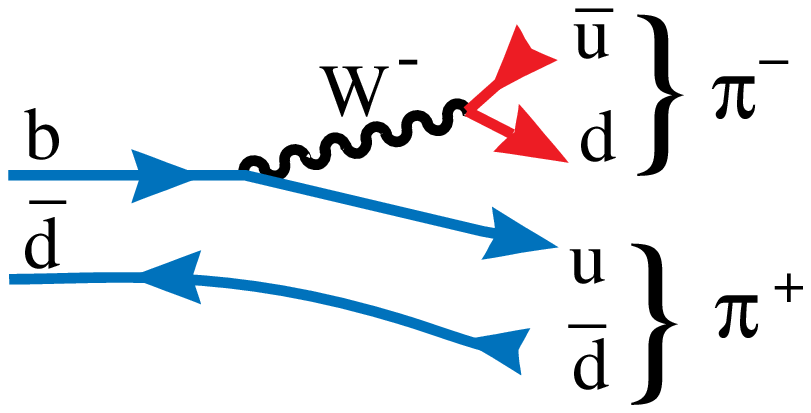,height=1.5in}
\epsfig{figure=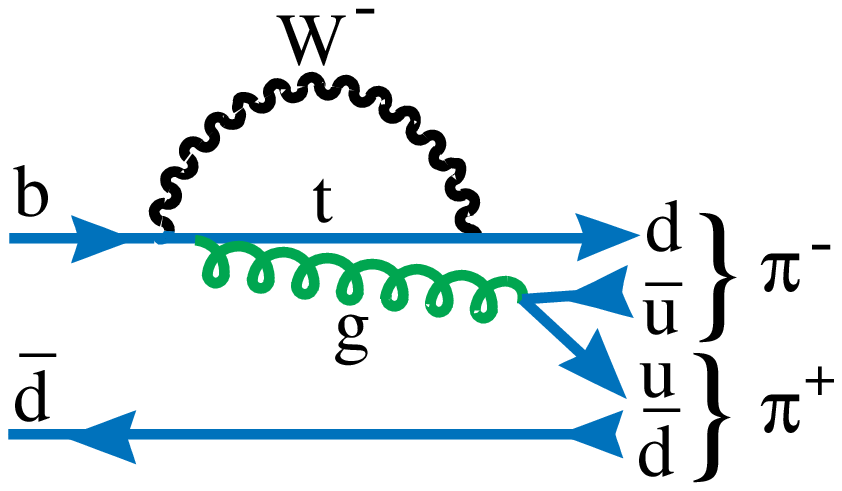,height=1.5in}}
\vspace{-.3cm}
\caption{\label{pippim}Decay diagrams for $\overline{B}^o\to\pi^+\pi^-$. (left) Via
tree level $V_{ub}$ moderated decay. (right) Via a Penguin process.}
%\vspace{-2mm}
\end{figure}

It is well known that $\sin (2\beta)$ can be measured without
problems caused by Penguin processes using the reaction $B^o\to J/\psi K_s$.
The simplest reaction that can be used to measure $\sin (2\alpha)$ is
$B^o\to \pi^+\pi^-$. This reaction can proceed via both the Tree and Penguin
diagrams shown in Fig.~\ref{pippim}.

Current measurements show a large Penguin component.
The ratio of Penguin {\it amplitude} to Tree {\it amplitude} in the
$\pi^+\pi^-$ channel is about 15\% in magnitude.
Thus the effect of the Penguin must be
determined in order to extract $\alpha$. The only model independent way 
of doing this was suggested by Gronau and London, but requires the measurement
of $B^{\mp}\to\pi^{\mp}\pi^o$ and $B^o\to\pi^o\pi^o$, the latter being rather 
daunting.

There is however, a theoretically clean method to determine $\alpha$.
The interference between Tree and Penguin diagrams can be exploited by
 measuring the time dependent CP violating
 effects in the decays $B^o\to\rho\pi\to\pi^+\pi^-\pi^o$  
as shown by Snyder and Quinn \cite{SQ}.

\begin{figure}[htb]
\vspace{-0.4cm}
\centerline{\epsfig{figure=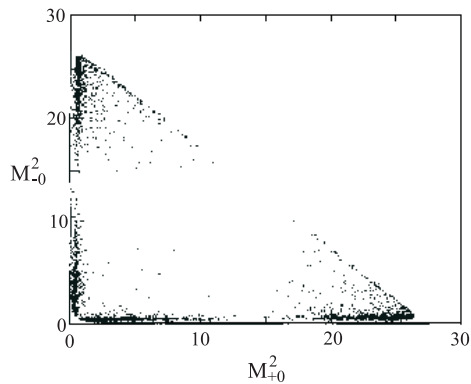,height=2.3in}}
\vspace{-.6cm}
\caption{\label{dalitz} The Dalitz plot for $B^o\to\rho\pi\to\pi^+\pi^-\pi^o$
from Snyder and Quinn.}
\end{figure}

The $\rho\pi$ final state has many advantages. First of all,
it has been seen with a relatively large rate. The 
branching ratio for the $\rho^o\pi^+$ final state as measured by CLEO is 
$(1.5\pm 0.5\pm 0.4)\times 10^{-5}$, and the rate for the neutral
 $B$ final state $\rho^{\pm}\pi^{\mp}$ is  
$(3.5^{+1.1}_{-1.0}\pm 0.5)\times 10^{-5}$, while the $\rho^o\pi^o$ final
state is limited at 90\% confidence level to $<5.1 \times 10^{-6}$
\cite{CLEO_rhopi}. (BABAR \cite{Bona01} measures
${\cal{B}}\left(B^o\to\rho^{\pm}\pi^{\mp}\right)$ as  
$(28.9\pm 5.4\pm 4.3)\times 10^{-6}$.) These
measurements are consistent with some theoretical expectations \cite{ali_rhopi}.
Furthermore, the associated vector-pseudoscalar
Penguin decay modes have conquerable or smaller branching ratios. Secondly, since the 
$\rho$ is spin-1, the $\pi$ spin-0 and the initial $B$ also spinless, the $\rho$ 
is fully polarized in the (1,0) configuration, so it decays as $\cos^2\theta$, 
where $\theta$ is the angle of one of the $\rho$ decay products with the other
$\pi$ 
in the $\rho$ rest frame. This causes the periphery of the Dalitz plot to be 
heavily populated, especially the corners. A sample Dalitz plot is shown in 
Fig.~\ref{dalitz}. This kind of distribution is good for maximizing the interferences, which 
helps minimize the error. Furthermore, little information is lost by excluding 
the Dalitz plot interior, a good way to reduce backgrounds.

To estimate the required number of events Snyder and 
Quinn preformed an idealized analysis that showed that a background-free,
flavor-tagged sample of 1000 to
2000 events was sufficient. The 1000 event sample usually yields good results 
for $\alpha$, but sometimes does not resolve the ambiguity. With the 2000 event sample, however, they always succeeded. 

This technique not only finds $\sin(2\alpha)$, it also determines 
 $\cos(2\alpha)$, thereby removing two of the remaining ambiguities. The final
 ambiguity can be removed using the CP asymmetry in $B^o\to\pi^+\pi^-$ and
 a theoretical assumption \cite{gross_quinn}.

\begin{figure}[htb]
\centerline{\epsfig{figure=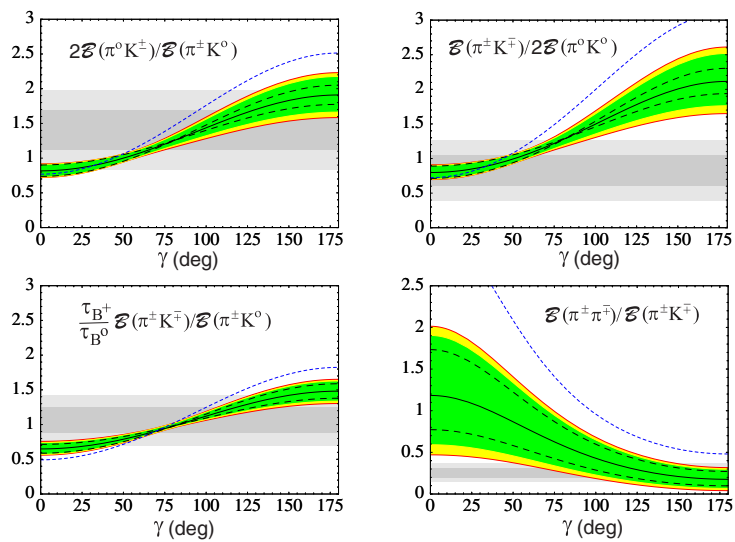,height=2.8in}}
\caption{\label{beneke1} Model predictions from Beneke \etal ~as
a function of the indicated rate ratios. The dotted curve shows the predictions
from naive factorization. The curved bands show the total model uncertainties
where the inner band comes from theoretical input uncertainties, while
the outer band allows for errors to corrections on the theory. The specific
sensitivity to $|V_{ub}|$ is showed as the dashed curves. 
The gray bands show the current data
with a $1\sigma$ error while the lighter bands are at $2\sigma$.}
\end{figure}

Several model dependent methods using the light two-body pseudoscalar decay
rates have been suggested for measuring $\gamma$ The basic idea in all these
methods can be summarized as follows: $B^o\to\pi^+\pi^-$ has the weak decay
phase $\gamma$. In order to reproduce the observed suppression of the decay
rate for $\pi^+\pi^-$ relative to $K^{\pm}\pi^{\mp}$ we require a large
negative interference between the Tree and Penguin amplitudes. This puts
$\gamma$ in the range of 90$^{\circ}$. There is a great deal of theoretical
work required to understand rescattering, form-factors etc... We are left with
several ways of obtaining model dependent limits, due to Fleischer and Mannel \cite{FM},
Neubert and Rosner \cite{NR},
Fleischer and Buras \cite{FB}, and Beneke \etal ~\cite{Beneke}. The latter 
make a sophisiticated model of QCD factorization and apply corrections.
Fig.~\ref{beneke1} shows values of $\gamma$
that can be found in their framework, once better data are obtainable.

In fact, it may be easier to measure $\gamma$ than $\alpha$ in a model independent manner. There have been two methods suggested.

\begin{figure}[htb]
\centerline{\epsfig{figure=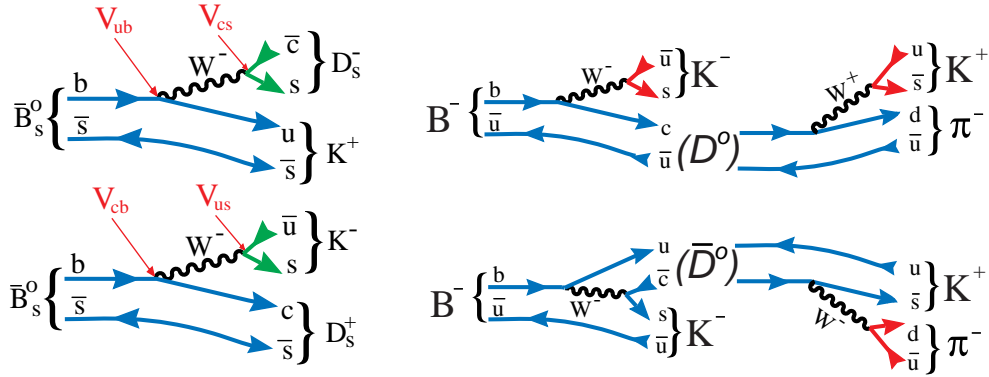,height=2in}}
\caption{\label{DK_both} (left) The two diagram diagrams for $B_s\to
D_s^{\pm}K^{\mp}$ that interfere via $B_s$ mixing. (right) The two interfering
decay
diagrams for $B^-\to \overline{D}^o K^-$ where one is a $b\to u$ transition and
the other a doubly-Cabibbo suppressed decay.}
\end{figure}

(1) Time dependent flavor tagged analysis of $B_s\to D_s^{\pm}K^{\mp}$. This
is a direct model independent measurement \cite{Aleks}. Here the Cabibbo
suppressed $V_{ub}$ decay interferes with a somewhat
less suppressed $V_{cb}$ decay via $B_s$ mixing as illustrated in Fig.~\ref{DK_both}
(left). Even though we are not dealing with CP eigenstates here there are no
hadronic uncertainties, though there are ambiguities.

(2) Measure the rate differences between $B^-\to \overline{D}^o K^-$ and
$B^+\to {D}^o K^+$ in two different $D^o$ decay modes such as $K^-\pi^+$
and $K^+ K^-$. This method makes use of the interference between the tree
and doubly-Cabibbo suppressed decays of the $D^o$, and does not depend
on any theoretical modeling \cite{sad}\cite{gronau}. See Fig.~\ref{DK_both}
(right).

\subsection{New Physics Tests in Specific Models}
\subsection{Supersymmetry}
Supersymmetry is a kind of super-model. The basic idea is that for every
fundamental fermion there is a companion boson and for every boson there
is a companion fermion. There are many different
implementations of couplings in this framework \cite{Masiero00}. In the most general case we pick up
80 new constants and 43 new phases. This is clearly too many to handle so
we can try to see things in terms of simpler implementations. In the minimum
model (MSSM) we have only two new fundamental phases. One, $\theta_D$, would
arise in $B^o$ mixing and the other, $\theta_A$, would appear in $B^o$ decay.
A combination would generate CP violation in $D^o$ mixing, call it $\phi_{K\pi}$
when the $D^o\to K^-\pi^+$ \cite{Nir}. Table~\ref{tab:MSSM} shows the CP
asymmetry in three different processes in the Standard Model and the MSSM.

\begin{table}[htb]
\vspace{-2mm}
\begin{center}
\caption{CP Violating Asymmetries in the Standard Model and the MSSM.
\label{tab:MSSM}}
\vspace*{2mm}
\begin{tabular}{lcl}\hline\hline
%\multicolumn{2}{c}{$\BM$}&\multicolumn{2}{|c}{$\BP$}
Process & Standard Model & New Physics\\
\hline
$B^o\to J/\psi K_s$ & $\sin 2\beta$ & $\sin2(\beta+\theta_D)$\\
$B^o\to \phi K_s$ & $\sin 2\beta$ & $\sin2(\beta+\theta_D+\theta_A)$\\
$D^o\to K^-\pi^+$ & 0 & $\sim \sin\phi_{K\pi}$ \\
\hline\hline
\end{tabular}
\end{center}
\end{table}
Two direct effects of New Physics are clear here. First of all, the difference in
CP asymmetries between $B^o\to J/\psi K_s$ and $B^o\to \phi K_s$ would show
the phase $\phi_A$. Secondly, there would be finite CP violation in 
$D^o\to K^-\pi^+$ where none is expected in the Standard Model.

Manifestations of specific SUSY models lead to different patterns.
Table~\ref{tab:fNir} shows the expectations for some of these models in
terms of these variables and the neutron electric dipole moment $d_N$;
see \cite{Nir} for details.
\begin{table}[htb]
\vspace{-2mm}
\begin{center}
\caption{Some SUSY Predictions.
\label{tab:fNir}}
\vspace*{2mm}
\begin{tabular}{lcccc}\hline\hline
%\multicolumn{2}{c}{$\BM$}&\multicolumn{2}{|c}{$\BP$}
Model & $d_N\times 10^{-25}$ & $\theta_D $& $\theta_A$ &$\sin\phi_{K\pi}$ \\
\hline
Standard  Model & $\leq 10^{-6}$ & 0 & 0 & 0\\
Approx. Universality & $\geq 10^{-2}$  & $\cal O$(0.2) & $\cal O$(1)  & 0\\
Alignment  & $\geq 10^{-3}$  & $\cal O$(0.2) & $\cal O$(1)  & $\cal O$(1)\\
Heavy squarks  & $\sim 10^{-1}$  & $\cal O$(1) & $\cal O$(1)  & $\cal
O$($10^{-2}$)\\
Approx. CP & $\sim 10^{-1}$  & -$\beta$ & 0  & $\cal O$($10^{-3}$) \\
\hline\hline
\end{tabular}
\end{center}
\end{table}
Note, that ``Approximate CP" has already been ruled out by the measurements
of $\sin 2\beta$.

\begin{figure}[htb]
\centerline{\epsfig{figure=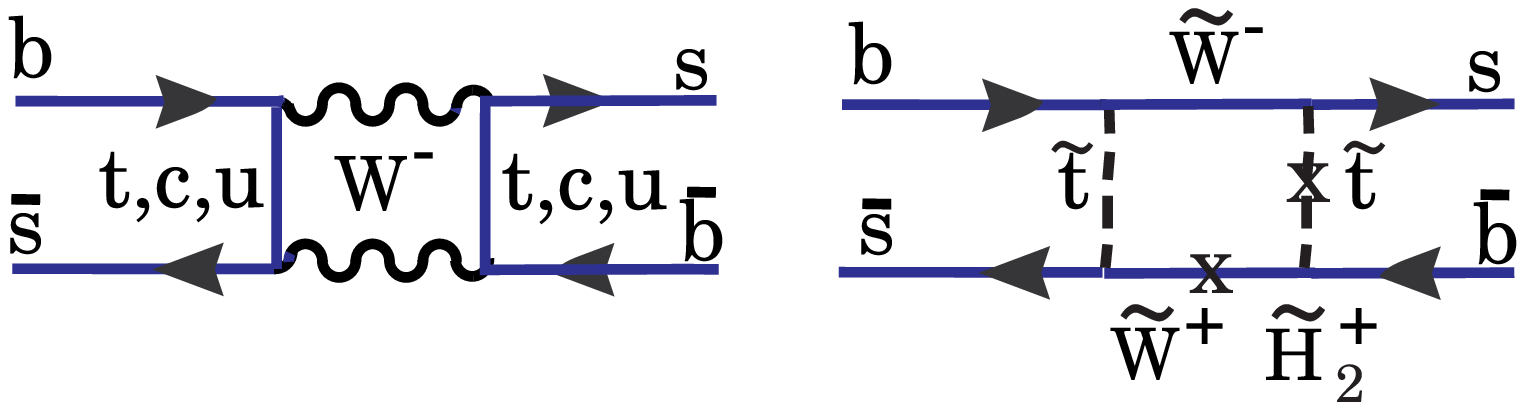,height=1.5in}}
\caption{\label{Bs_mssm} The Standard Model (left) and MSSM (right)
contributions to $B_s^o$ mixing.}
\end{figure}

In the context of the MSSM there will be significant contributions to $B_s$
mixing, and the CP asymmetry in the charged decay $B^{\mp}\to \phi K^{\mp}$.
The contribution to $B_s$ mixing significantly enhances the CP violating
asymmetry in modes such as $B_s\to J/\psi \eta$. (Recall the CP asymmetry in
this mode is proportional to $\sin2\chi$ in the Standard Model.) 
The Standard Model and MSSM diagrams are shown in 
Fig.~\ref{Bs_mssm}. The expected CP asymmetry in the MSSM is
$\approx \sin\phi_{\mu}\cos\phi_A\sin(\Delta m_s t)$, which is approximately
10 times the expected value in the Standard Model \cite{Hinch01a}.

\begin{figure}[htb]
\centerline{\epsfig{figure=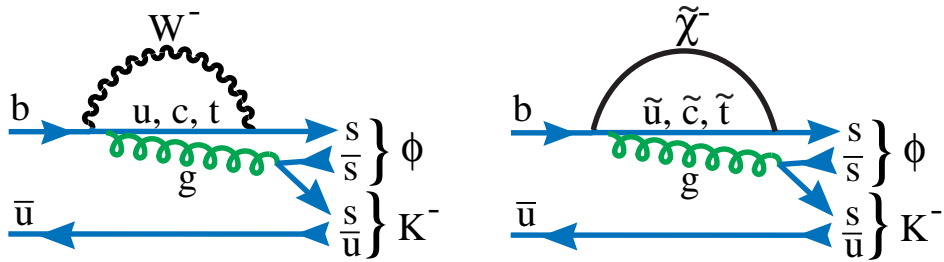,height=1.5in}}
\caption{\label{Phi_K_mssm} The Standard Model (left) and MSSM (right)
contributions to $B^-\to \phi K^-$.}
\end{figure}

We observed that a difference between CP asymmetries in 
$B^o\to J/\psi K_s$ and $\phi K_s$ arises in the MSSM due to a CP asymmetry in
the decay phase. It is possible to observe this directly by looking for
a CP asymmetry in $B^{\mp}\to \phi K^{\mp}$. The Standard Model and MSSM
diagrams are shown in Fig.~\ref{Phi_K_mssm}. Here the interference of the two
diagrams provides the CP asymmetry. The predicted asymmetry is equal to
$\left(M_W/m_{squark}\right)^2\sin\phi_{\mu}$ in the MSSM, where $m_{squark}$
is the relevant squark mass \cite{Hinch01a}. 

The $\phi K$ and $\phi K^*$ final states have been observed, first by CLEO 
\cite{Briere01} and
subsequently by BABAR \cite{Aubert01}.  The average branching ratio is 
${\cal B}(B^-\to\phi K^-)=(6.8\pm 1.3)\times 10^{-6}$ showing that in principle
large samples can be acquired especially at hadronic machines.

\subsection{Other New Physics Models}
There are many other specific models that predict New Physics in $b$ decays.
I list here a few of these with a woefully incomplete list of references, to
give a flavor of what these models predict. 
\begin{itemize}
\item {\it Two Higgs and Multi-Higgs Doublet Models-} They predict large
effects in $\epsilon_K$ and CP violation in $D^o\to K^-\pi^+$ with
only a few percent effect in $B^o$ \cite{Nir}. Expect to see 1-10\% CP
violating effects in $b\to s\gamma$ \cite{Wolfenstein94}.
\item {\it Left-Right Symmetric Model-} Contributions compete with or even
dominate over Standard Model contributions to $B_d$ and $B_s$ mxing. This means
that CP asymmetries into CP eigenstates could be substantially different from
the Standard Model prediction \cite{Nir}.
\item {\it Extra Down Singlet Quarks-} Dramatic deviations from Standard Model
predictions for CP asymmetries in $b$ decays are not unlikely \cite{Nir}.
\item {\it FCNC Couplings of the $Z$ boson-} Both the sign and magnitude of the
decay leptons in $B\to K^*\ell^+\ell^-$ carry sensitive information on new
physics. Potential effects are on the of 10\% compared to an entirely
negligable Standard Model asymmetry of $\sim 10^{-3}$ \cite{Buchalla00}. These
models also predict a factor of 20 enhancement of $b\to d\ell^+\ell^-$ and 
could explain a low value of $\sin2\beta$ \cite{Barenboim01a}.
\item {\it Noncommutative Geometry-} If the geometry of space time is
noncommutative, i.e. $[x_{\mu},x_{\nu}]=i\theta_{\mu\nu}$, then CP violating
effects may be manifest a low energy. For a scale $<$2 TeV there are comparable
effects to the Standard Model \cite{Hinch01b}.
\item {\it MSSM without new flavor structure-} Can lead to CP violation in
$b\to s\gamma$ of up to 5\% \cite{Bartl01}. Ali and London propose \cite{Ali99}
that the Standard Model formulas are modified by Supersymmetry as
\begin{eqnarray}
\Delta m_d &=&\Delta m_d{\rm(SM)}\left[1+f\left(m_{\chi^{\pm}_2},m_{\tilde{t}_R},
m_{H^{\pm}},tan\beta\right)\right] \\
\Delta m_s &=&\Delta m_s{\rm(SM)}\left[1+f\left(m_{\chi^{\pm}_2},m_{\tilde{t}_R},
m_{H^{\pm}},tan\beta\right)\right] \\
\left|\epsilon_K\right|&=&{G_F^2f^2_KM_KM_W^2\over 6\sqrt{2}\pi^2\Delta M_K}
B_K(A^2\lambda^6\overline{\eta})\left[y_c\left(\eta_{ct}f_3(y_c,y_t)-\eta_{cc}
\right)\right. \nonumber \\
& &\left.+\eta_{tt}y_tf_s(y_t)\left[1+f\left(m_{\chi^{\pm}_2},m_{\tilde{t}_R},
m_{H^{\pm}},tan\beta\right)\right]A^2\lambda^4(1-\overline{\rho})\right]~~,
\end{eqnarray}
where $\Delta m (SM)$ refers to the Standard Model formula and the expression
for $\left|\epsilon_K\right|$ would be the Standard Model expression if $f$
were set equal to zero. Ali and London show that it is reasonable to expect
that $0.8>f>0.2$ so since the CP violating angles will not change
from the Standard Model, determining the value of $(\rho,~\eta)$ using the
magnitudes $\Delta m_s/\Delta m_d$ and $|\epsilon_K|$ will show an
inconsistency with values obtained using other magnitudes and angles. 
\item {\it Extra Dimensions-} We are beginning to see le to expect
that $0.8>f>0.2$ so since the CP violating angles will not change
from the Standard Model, determing the value of $(\rho,~\eta)$ using the
magnitudes $\Delta m_s/\Delta m_d$ and $|\epsilon_K|$ will papers predicting $b$ decay
phenomena when the world has extra dimensions. See \cite{Agashe01}.
\end{itemize}

I close this section with a quote from  Masiero and Vives \cite{Masiero01}:
``The relevance of SUSY searches in rare processes is not confined to the
usually quoted possibility that indirect searches can arrive `first' in signaling
the presence of SUSY. Even after the possible direct observation of SUSY
particles, the importance of FCNC and CP violation in testing SUSY remains of
utmost relevance. They are and will be complementary to the Tevatron and LHC
establishing low energy supersymmetry as the response to the electroweak
breaking puzzle."

I agree, except that I would replace ``SUSY" with ``New Physics."

%fig1
%\begin{figure}
%[htbp]
%\centering
%\rule{2mm}{22mm} I use this whilst editing without figs
%\includegraphics[width=6cm,clip,trim=0 0 40 40]{fig1.eps}
%\caption{The cluster is treated as a
%point mass $M_c$ in uniform circular motion of angular velocity
%$\omega$ at a distance $R$ from a point-mass galaxy $M_g$.}
%\label{derivation}
%\end{figure}

\section{Conclusions}

It is clear that precision studies of $b$ decays can bring a wealth of
information to bear on new physics, that probably will be crucial in sorting
out anything seen at the LHC. This is possible because we do expect to have
data samples large enough to test these ideas from
existing and approved experiments. In Table~\ref{tab:expectations} I show the expected rates in
BTeV for one year of running ($10^7$ s) and an $e^+e^-$ $B$-factory operating
at the $\Upsilon (4S)$ with a total accumulated sample of 500 fb$^{-1}$, about
what is expected around 2006. (LHCb numbers are the same order of magnitude
as the BTeV numbers for many of the modes.) 

\section*{Acknowledgements}
This work was supported by the U. S. National Science Foundation. 
 My colleagues at BTeV and CLEO contributed much to my understanding. In
particular I thank, M. Artuso, J. Butler and T. Skwarnicki.
\begin{table}[htb]
\vspace{-2mm}
\begin{center}
\caption{Comparison of BTeV and $B$-factory Yields on Different Time Scales.
\label{tab:expectations}}
\vspace*{2mm}
\begin{tabular}{l|rrrrrr}\hline\hline
Mode&\multicolumn{3}{c}{BTeV $(10^7s)$}& \multicolumn{3}{c}{$B$-factory (500
fb$^{-1}$)}\\
  & Yield & Tagged$^{\dagger}$ & S/B & Yield & Tagged$^{\dagger}$ & S/B\\
\hline
$B_s\to J/\psi\eta^{(')}$ & 22000 & 2200 &$>$15 & &&\\
$B^-\to\phi K^-$ & 11000 & 11000 & $>$10 & 700 & 700 &4\\
$B^o\to\phi K_s$ & 2000 & 200 & 5.2 & 250 & 75 & 4\\
$B^o\to K^{*o}\mu^+\mu^-$ & 4400 &4400 & 11 & $\sim$50 & $\sim$50 & ?\\
$D^{*+}\to\pi^+ D^o$; $D^o\to K^-\pi^+$ & $\sim 10^8$ &$\sim 10^8$ & large &
$8\times 10^5$ & $8\times 10^5$ & large \\\hline\hline
\multicolumn{7}{l}{$\dagger$ Tagged here means that the intial flavor of the $B$
is determined.} 
\end{tabular}

\end{center}
\end{table}

\end{document}